# Flexoelectric fracture-filter effect in ferroelectrics

Kumara Cordero-Edwards[1*⌇], Hoda Kianirad[2], Jordi Sort[3,4], Carlota Canalias[2], and Gustau Catalan[1,4]*

[1] Catalan Institute of Nanoscience and Nanotechnology (ICN2), CSIC and BIST, Campus UAB, Bellaterra, Barcelona 08193, Catalonia.

[2] Department of Applied Physics, KTH-Royal Institute of Technology, Roslagstullsbacken 21, 10691, Stockholm, Sweden

[3] Departament de Física, Universitat Autònoma de Barcelona (UAB), Edifici Cc, E-08193 Bellaterra, Spain

[4] Institució Catalana de Recerca i Estudis Avançats (ICREA), Pg. Lluís Companys 23, E-08010 Barcelona, Catalonia.

* Corresponding author:

Email: rohini.corderoeduards@unige.ch    Tel: + 41 022 379 30 34

Email: gustau.catalan@icn2.cat    Tel: + 34 93 737 3618

⌇ Now at University of Geneva

**Abstract**

The propagation front of a crack generates large strain gradients and it is therefore a strong source of gradient-induced polarization (flexoelectricity). Herein, we demonstrate that, in piezoelectric materials, a consequence of flexoelectricity is that crack propagation will be helped or hindered depending on whether it is parallel or antiparallel to the piezoelectric polar axis. This means that the theory of fracture physics can no longer assume mechanical symmetry in polar materials. The discovery of fracture asymmetry also has practical repercussions for the electromechanical fatigue of ferroelectrics and piezoelectric transducers, as well enabling a new degree of freedom for crack-based nanopatterning.

**Introduction**

Crack propagation causes materials to break, and forms basis of fracture physics -a vital element of materials science and device engineering. Controlled cracking has also been proposed as a new mechanism for device nano-patterning,[1] turning the harnessing of crack propagation into a constructive pursuit. In piezoelectric and ferroelectric materials, fracture physics is additionally important because voltage-induced strains cause the appearance and propagation of microcracks that result in material fatigue and ultimate failure of piezoelectric transducers.[2-3] Understanding the fracture behaviour of piezoelectrics is therefore very important. Here we show that crack-generated flexoelectricity causes in ferroelectrics an original and hitherto unnoticed valve-like or "crack filter" behaviour, whereby crack propagation is facilitated or impaired depending on the sign of the ferroelectric polarization.

Flexoelectricity [4,5,6] has disruptive consequences for the functional and mechanical properties of materials [7,8,9,10]. For example, the flexoelectric fields generated by cracks have been shown to be strong enough to be able to trigger the self-repair process in bone fractures.[11] In the case of ferroelectrics, flexoelectricity enables qualitatively new behaviours. For example, it has recently been predicted [10] and demonstrated [12] that ferroelectrics can have an asymmetric mechanical response to inhomogeneous deformations. Because fracture fronts concentrate the biggest local deformations that a solid can withstand, flexoelectricity is also expected to affect the fracture physics of ferroelectrics.[10] The present work demonstrates a fundamentally new fracture phenomenon due flexoelectricity: is that crack propagation in ferroelectrics is asymmetric and switchable, so that cracks propagating parallel to the ferroelectric polarization become longer than those traveling against it.

In the present experiment, Vickers Indentation Tests were performed on a Rb-doped KTiOPO$_4$ (RKTP) single crystal with the polarization in-plane. We chose this ferroelectric because it is uniaxial, and thus ferroelastic effects can be excluded. RKTP is also technologically relevant material, commonly used as a frequency conversion device in nonlinear optics. [13-14] For such applications, a bulk periodic domain pattern with alternating domain orientations (periodic poling) is created in the crystal. The procedure for this is well-established [13], and therefore poling the crystal was possible. Poling of antiparallel domains on the same crystal was used in order to ensure that geometrical effects such as a slight tilt or miscut of the crystal surface did not affect the results. By poling two domains of antiparallel orientation, indents could be performed on domains of opposite polarities on the same surface in the same experiment, as sketched in figure 1a. That way, the effect of alternating polarity could be tested without affecting any other geometrical parameter or changing the sample.

Mechanical tests were conducted applying sets of 200mN and 300mN loads, with the orientation of the indenter being such that two of its four corners were parallel to the polar axis and the other two perpendicular. In order to control for statistical fluctuations in fracture toughness, 30 indents for each force (15 for each domain polarity) were performed, with each indent generating four cracks along the parallel, antiparallel and perpendicular directions. In total, 240 cracks were hence analysed. The radial crack lengths, from the corners of the indents (see inset in Figure 1a), were measured with an optical microscope immediately after indentation. A sketch of the experiment is in Figure 1(a), and two indentation samples can be seen in Figure 1(b) and (c).

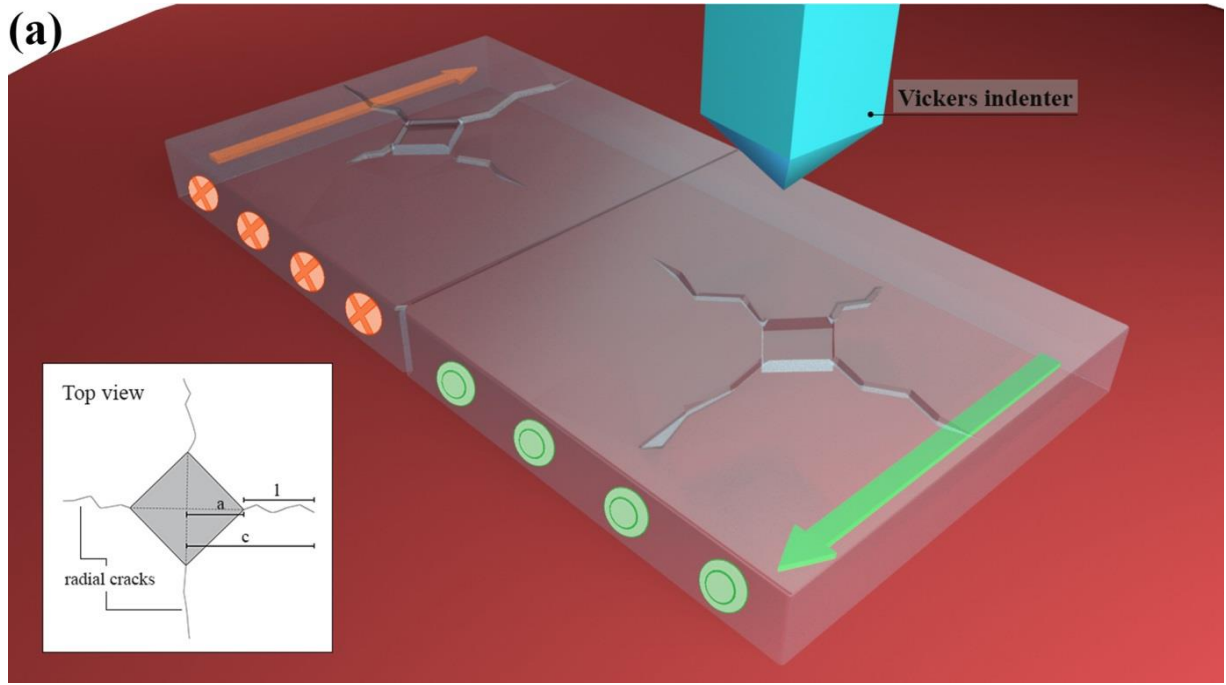

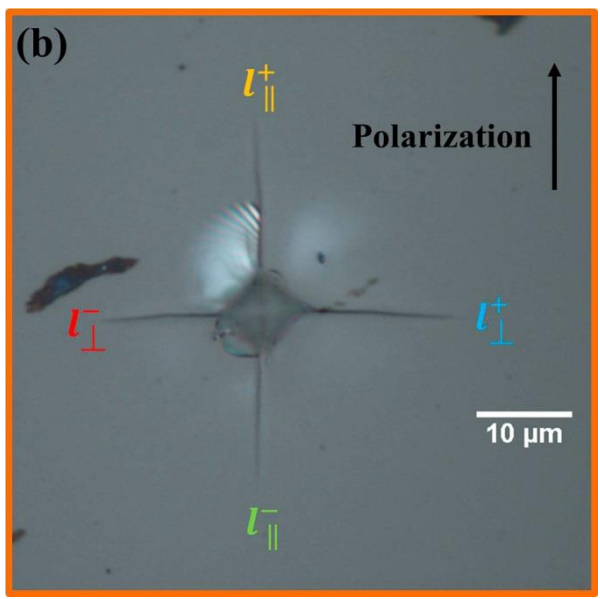
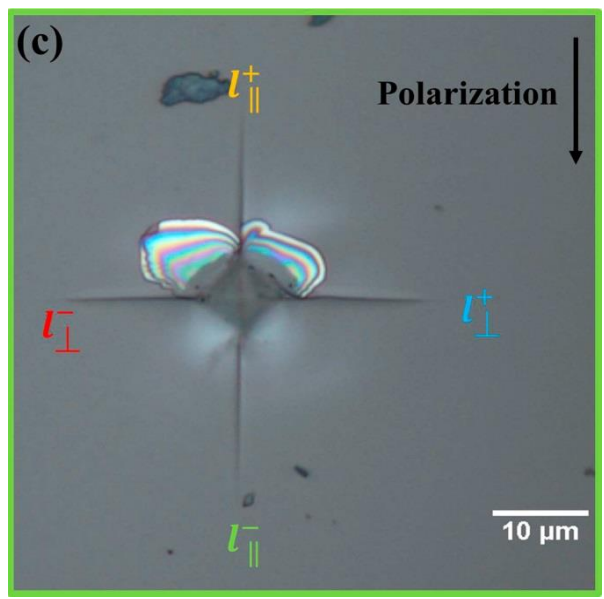

Figure 1: (a) Schematic of the Vickers Indentation test showing the top view of typical radial crack propagation for indentation fracture toughness measurement with corresponding crack (*l*) and diagonal lengths (*2a*). Optical micrographs of Vickers indent in RKTP showing the radial crack propagation for (b) up and (c) down polarization.

**Results**

After measuring the length of the cracks ($l$), the length asymmetry along the polar axis was calculated for each indentation. To verify that the results were not artefacts, we also measured the asymmetry in the direction perpendicular to the polar direction, where in theory there should be none. We define the asymmetry coefficient as

$$\%Asy = \frac{l^+ - l^-}{\langle l \rangle} * 100, \qquad (1)$$

where $l^+$ is the crack length parallel to the polarization, and $l^-$ is the crack length antiparallel to the polarization (up or down in the plan-view photos). For cracks perpendicular to the poling direction, + and − designate right or the left directions, respectively, in the plan-view photos. The average crack length is $\langle l \rangle \equiv \frac{l^+ + l^-}{2}$. Positive (negative) asymmetry indicates a longer (shorter) crack than the average. When cracks have the same length, the asymmetry coefficient is zero.

Figure 2(a) shows the asymmetry of the cracks perpendicular to the polar axis. For these, as expected, there is no asymmetry within statistical error. This lack of perpendicular asymmetry provides a safety check for the robustness of the experiment and a background for comparison. In contrast to the perpendicular cracks, Figure 2(b) shows that cracks parallel to the poling direction are asymmetric: for $P^+$ domains, a positive asymmetry is measured, and the asymmetry is reversed for the $P^-$ domains. In other words: crack length parallel to the polarization is always greater than crack length antiparallel to the polarization, irrespective of the polarity of the domain.

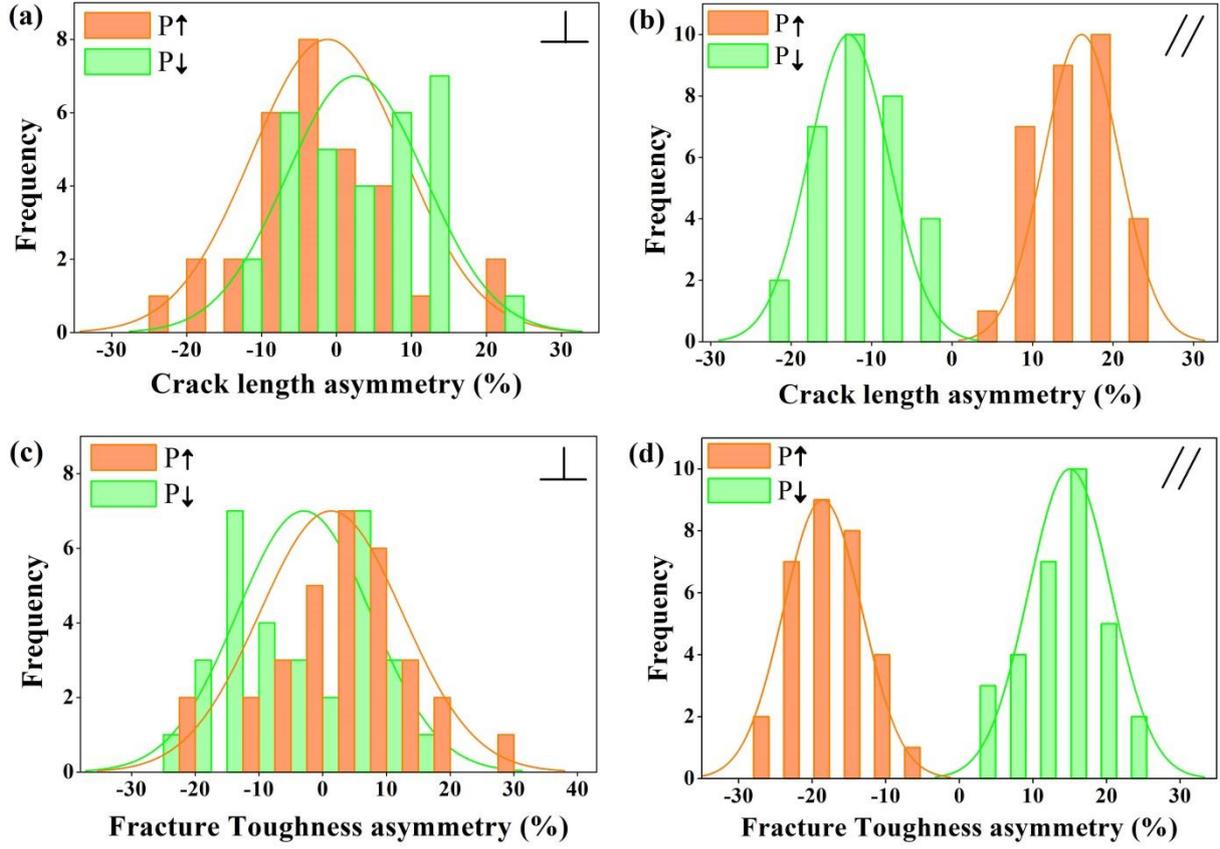

Figure 2: Crack length asymmetry (a) perpendicular and (b) parallel to the polar axis. Fracture toughness asymmetry (c) perpendicular and (d) parallel to the polar axis.

The asymmetry of crack length can be used to quantify the asymmetry in fracture toughness, which measures the stress intensity required for creating a crack [15]. Fracture toughness is given by [16]

$$K_{IC} = 0.016 * \left(\frac{E}{H}\right)^{1/2} \left(\frac{F}{c^{3/2}}\right), \quad H = \frac{F}{2a^2}, \tag{2}$$

where $E$ is the Young Modulus, $H$ the Vickers hardness, $F$ the indent load, $c$ is the distance from the center of the indentation impression to the tip of the crack, and $2a$ is the diagonal of the indent (see inset in Figure 1a). Using the values obtained from our tests, $K_{IC}$ was obtained for each crack, and then using the expression (1), the asymmetries were calculated.

Figures 2(c) and 2(d) show the asymmetry for the perpendicular and parallel direction, respectively. As expected, there is asymmetry only along the polar axis, i.e. when ferroelectric and flexoelectric polarizations are parallel (crack propagating in the same direction as the ferroelectric polarization), or antiparallel (crack propagating in the opposite direction as the ferroelectric polarization), and no asymmetry in the perpendicular direction. The average value of the fracture toughness for cracks parallel to the polarization was ~ 0.19 ± 0.02 MPam$^{1/2}$, whereas for the ones antiparallel to the polarization it was ~ 0.23 ± 0.03 MPam$^{1/2}$. In other words, in ferroelectric RKTP, fracture toughness is enhanced (yielding to shorter cracks) by 20% when flexoelectricity and ferroelectricity are antiparallel compared to when they are parallel.

Since all indentations are performed under the exact same geometrical conditions; the fact that the asymmetry is reversed when the polarization is inverted implies that the origin of the asymmetry is not a geometrical artifact. Differences in surface adsorbates or near-surface defects can also be excluded; even if such differences did exist (and none should be expected given that the polarization is in-plane), each pair of cracks is generated in the same spot and thus encounters identical surface conditions. The asymmetry in crack propagation is therefore intrinsic and linked to polarity. Ferroelectric polarity acts as a partial "valve" that can be switched to facilitate or impair crack propagation.

The basis of the asymmetry is the interplay between flexoelectricity and ferroelectricity. [10, 12, 17] The local deformation at the tip of the crack generates a flexoelectric polarization[10]. The electrostatic energy cost of this flexoelectric polarization depends on whether it is parallel or antiparallel to the ferroelectric polarization, thus resulting in different energy costs for cracking: a higher electrostatic energy means a higher energy cost for crack propagation, and thus a shorter crack.

The above arguments should be valid for any piezoelectric material. For ferroelectrics, however, there is in theory an additional source of mechanical asymmetry, which is that the flexoelectric field near the tip of the crack may be large enough to cause local switching of the polarization [18-19], thus providing an additional path for energy dissipation that further reduces the available energy for mechanical fracture. This process is known as switching-induced toughening [20, 21, 22], and the size of the switching region can be calculated by comparing the electrostatic energy cost of switching (switched polarization multiplied by coercive field) against the mechanical and electromechanical energy provided by the crack[20]. Switching-induced toughening has so far been studied in ferroelastic-ferroelectrics (i.e. ferroelectric materials where mechanical stress can switch the easy axis direction), but flexoelectricity in principle also enables purely ferroelectric (180 degree) switching in non-ferroelastic uniaxial ferroelectrics. Here we examine the extent to which such effect can contribute to the observed cracking asymmetry of our samples.

Considering a uniaxial ferroelectric, and adding a flexoelectric term to the energy balance, switching should occur when:

$$f_{ijkl} \in_{j,kl} \Delta P_i + E_i \Delta P_i \geq 2 P_s E_c \qquad (3)$$

where $f_{ijkl}$ is the flexocoupling tensor, $\in_{j,kl}$ is the strain gradient, and $\Delta P_i$ are the changes in the spontaneous polarization during the switching,; $P_s$ is the magnitude of the spontaneous polarization; and $E_c$ the coercive electric field. Since there is no external electric field, we can discard the second term, and $\Delta P_i = 2P_s$ for 180° domain switching.[21] The condition for switching thus simplifies to $f_{ijkl} \in_{j,kl} \geq E_c$. In other words, switching happens when the flexoelectric field (left side of the equation) exceeds the coercive field (right side term).

To estimate the size of the switched region, we have simplified the problem by neglecting the shear strain gradient and considering only longitudinal and transverse components, assuming flexocoupling coefficients of the order of $f = 10V$, as generally observed for ceramics [6, 23]. With these simplifications, switching should occur in the region of the ferroelectric crystal that satisfies the condition:

$$\left(\frac{\partial \epsilon_{22}}{\partial x_1} + \frac{\partial \epsilon_{11}}{\partial x_1}\right) \geq \frac{E_c}{f} \quad (4)$$

Considering the coercive field of RKTP ($E_c = 3.7 \times 10^6$ Vm$^{-1}$), and with the aforementioned simplfications, a total strain gradient of ~$3.7 \times 10^5$ m$^{-1}$ is theoretically required to induce switching in RKTP. To see whether such strain gradients are reached in the vicinity of the crack, we have used elastic theory to calculate the strain field[24]

$$\varepsilon_{ij}^{el} = \frac{1+v}{E}\sigma_{ij} - 3\frac{v}{E}\sigma_m \delta_{ij}, \quad (5)$$

where $\sigma_{ij}$ is the stress applied to the crack in each direction, and its expression depends on the propagation modes; $\sigma_m$ is the average stress, $E$ is the Young's Modulus, and $v$ is the Poisson ratio. Focusing on crack mode I (tensile loading), the stress fields in this type of crack are given by the following equations,

$$\sigma_{11} = \frac{K_I}{\sqrt{2\pi r}} \cos\frac{\theta}{2} \left(1 - \sin\frac{\theta}{2} \sin\frac{3\theta}{2}\right) \quad (6)$$

$$\sigma_{22} = \frac{K_I}{\sqrt{2\pi r}} \cos\frac{\theta}{2} \left(1 + \sin\frac{\theta}{2} \sin\frac{3\theta}{2}\right) \quad (7)$$

$$\tau_{12} = \frac{K_I}{\sqrt{2\pi r}} \cos\frac{\theta}{2} \sin\frac{\theta}{2} \cos\frac{3\theta}{2}, \quad (8)$$

where $K_I$ is the intensity factor (fracture toughness for this calculation). Transforming equations (6), (7) and (8) to Cartesian coordinates, and using Mathematica[26] for the calculations, we have computed analytically the strain field in equation (5), and the strain gradient associated with it. The value used for the intensity factor (fracture toughness) was the one obtained in this study, $K_I$ = 0.23MPam$^{1/2}$; all other values were taken from the literature.[25]

The calculated 2D strain and strain gradient maps around a crack tip in RKTP are plotted in Figure 3(a) and 3(b) respectively. The dashed line outlines the region within which there is theoretically enough flexoelectricity to induce local switching of the polarization. We see that, as the crack propagates, it should switch the polarization in a volume of ~ 10 nm around the crack tip, thus dissipating energy and therefore reducing the maximum length that the crack can reach. The calculated size of this switching region, however is so small as to be beyond the resolution limit of PFM, plus it is also close to the limit for thermodynamic stability of a switched domain embedded in a non-switched matrix.[19, 27-28] Indeed, we examined the cracks by PFM finding no evidence of 180° local switching near them.

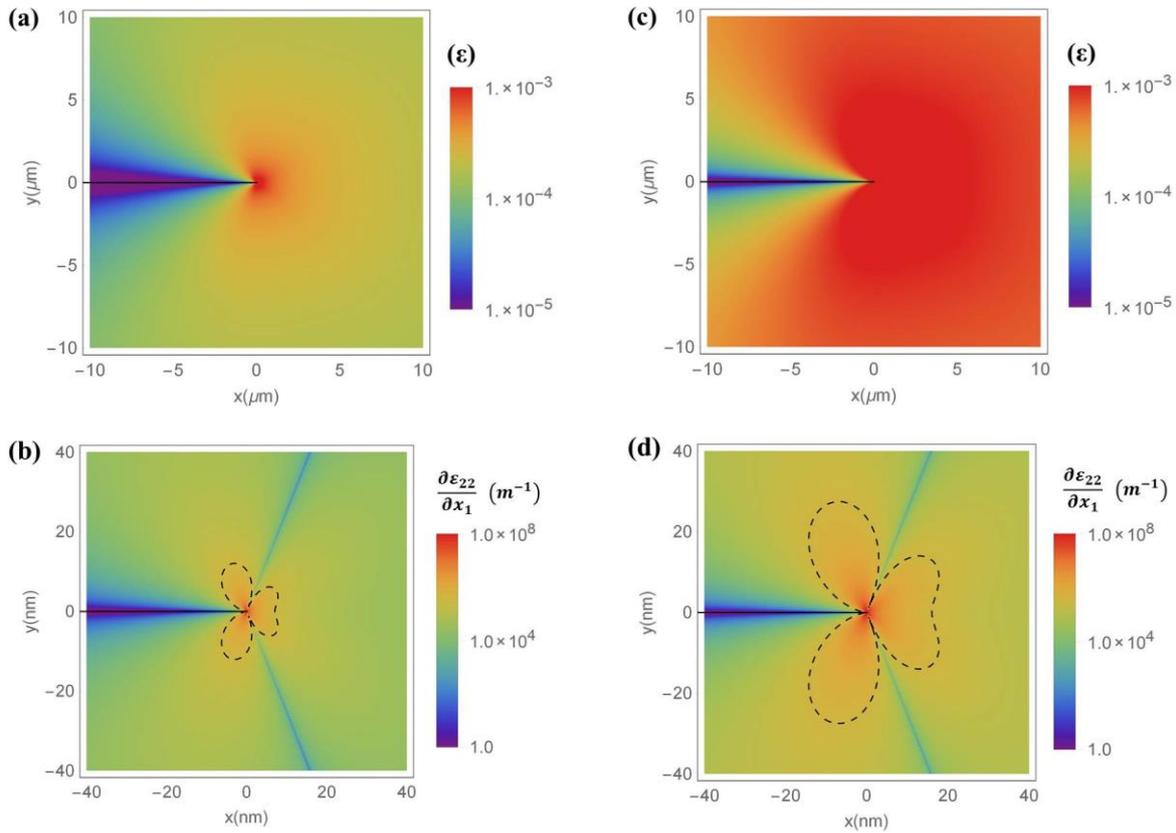

Figure 3: Calculated distribution of the (a) strain field, (b) strain gradient transversal component around the apex of a crack in RKTP. (c) Strain field, (d) strain gradient transversal component around the apex of a crack in LN. The black line marks the region where the gradient-induced electric field is strong enough to be able to induce local switching of the polarization

Flexoelectric effects are proportional to dielectric permittivity [29], therefore, it is to be expected that in a ferroelectric, with a higher dielectric constant, such as Lithium Niobate (LN) ($\kappa_{effRKTP} = 13$; $\kappa_{effLN} = 37$),[30] the local flexoelectric switching is enhanced. The coercive field for LN is $E_c = 2.1 \times 10^7$ Vm$^{-1}$ [31], by using the flexocoupling coefficient estimated elsewhere [12], we obtained that the strain gradient required to induce local switching is $4.2 \times 10^5$ m$^{-1}$. Using equations (5) – (8), and the values in the literature [32 - 34], we found that the distance from the crack tip that corresponds to this strain gradient is ~ 25 nm around the tip, which is big enough to be stable and detectable at room temperature. In order to confirm this experimentally, indentation tests were performed in LN y-cut, in the same conditions as with the RKTP sample, and afterward PFM images were taken.

The microscopy images of the resulting indent and cracks are shown in figure 4. In LNO, the easy fracture plane axis is not parallel to the polarization, but at 60 degrees from it, so the cracks generated at the corners of the indent tend to zig-zag instead of following a clean straight line along the polar axis. This makes it impossible to reliably measure and compare their lengths, but it does not affect their ability to generate flexoelectric fields. Indeed, the PFM images in Figure 4(b) and 4(c) show that cracks with a propagation component antiparallel to the polarization induce local 180 degree switching, leaving a trail of antiparallel domains in the crack's wake. LiNbO$_3$ is not ferroelastic and there is no external electric field, so the only explanation available for the observed 180 degree switching is the crack-generated flexoelectricity, in agreement with our calculations. The physics of this "flexoelectric switching" is ultimately the same that enables switching ferroelectric domains in thin films using AFM tip indentation [18]. The mechanical consequences, however, are profound: since switching dissipates energy, the cracks that switch polarization dissipate energy faster and thus cannot grow as long as those that do not. Consequently, fracture patterns in ferroelectrics must necessarily be asymmetric.

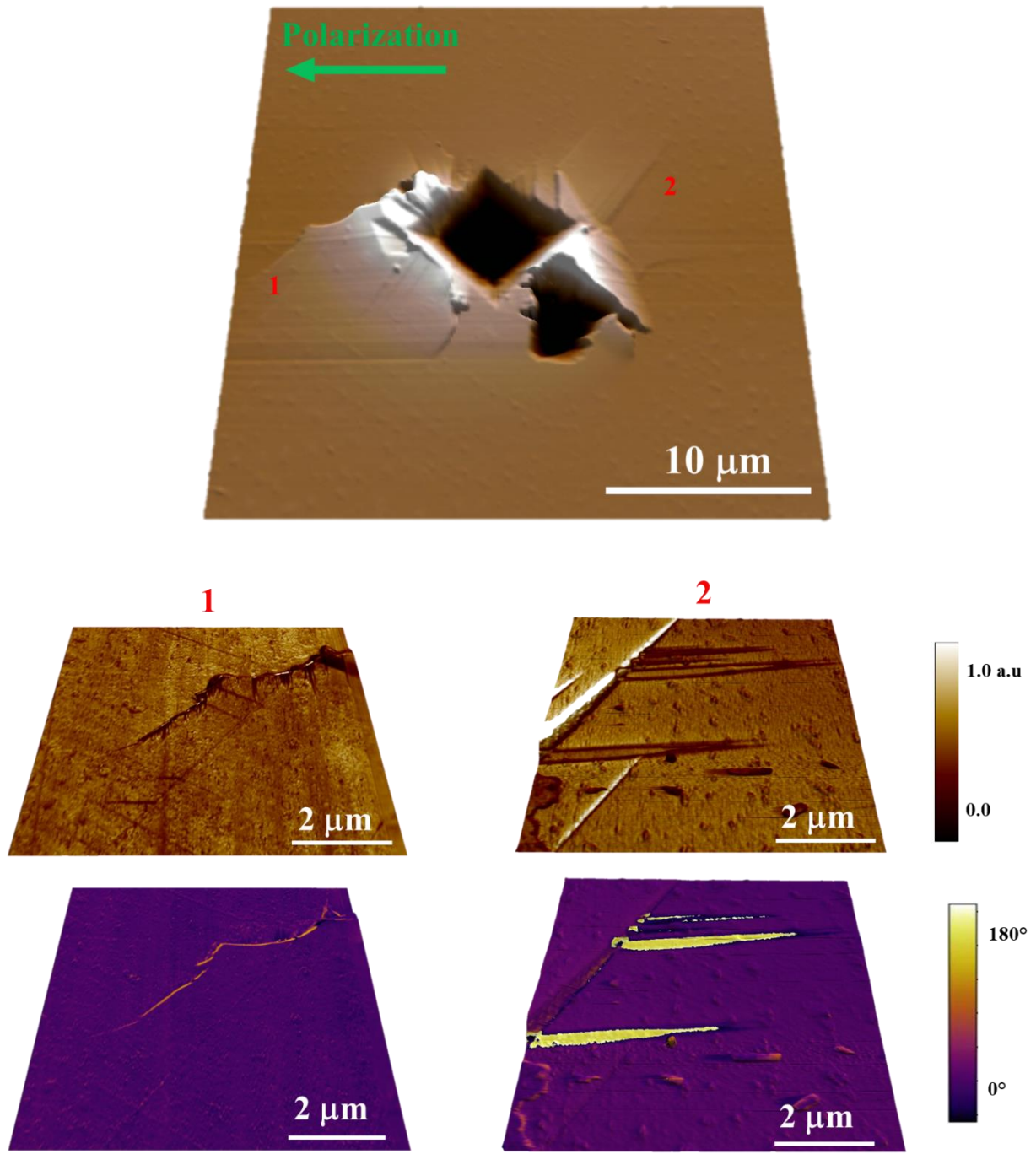

Figure 4: (top) AFM topography of Vickers indent in LN y-cut showing the radial crack propagation. (bottom) LPFM amplitude and phase showing local switching as the crack propagates opposite to the polarization of the crystal.

In summary, the interaction between flexoelectricity and ferroelectricity in fracture fronts leads to qualitatively new phenomena. F: first, cracks propagate more easily parallel to the polarization than antiparallel to it. This kind of "crack valve" behaviour means that inversion symmetry can no longer be assumed in the theoretical modelling of fractures in piezoelectric materials, and this is something that will have to be taken into account in any future modelling of fracture patterns in polar materials. The findings also have practical implications for fatigue in ferroelectrics and piezoelectric transducers; in particular, the results suggest that electromechanical fatigue due to microcracking could be enhanced or mitigated according to the poling direction of the ferroelectric transducer. Additionally, crack-diode functionality offers new degrees of freedom for crack-controlled nanopatterning,[1] as it suggests that polarity can be use to manipulate the fracture pattern. Finally, the observation that crack-induced flexoelectricity can cause ferroelectric switching adds a converse effect: it is not only that polarity affects crack propagation, but also that crack propagation can modify polarity.

**Acknowledgements**

K.C.-E. and G.C. acknowledge ERC Starting Grant 308023. ICN2 is supported by the Severo Ochoa program from Spanish MINECO (Grant No. SEV-2013-0295), and also Funded by the CERCA Programme / Generalitat de Catalunya. H.K. and C.C. acknowledge the Swedish Research Council for generous support. J.S work has been partially funded by the 2014-SGR-1015 project from the Generalitat de Catalunya and the MAT2014-57960-C3-1-R project from the Spanish Ministerio de Economía y Competitividad (MINECO), cofinanced by the 'Fondo Europeo de Desarrollo Regional (FEDER).